# Nonequilibrium Thermodynamics of Wealth Condensation

*Dieter Braun*

*Applied Physics,
Ludwig Maximilians University München,
Amalienstr. 54, D-80799 München, Germany*

We analyze wealth condensation for a wide class of stochastic economy models on the basis of the economic analog of thermodynamic potentials, termed transfer potentials. The economy model is based on three common transfers modes of wealth: random transfer, profit proportional to wealth and motivation of poor agents to work harder. The economies never reach steady state. Wealth condensation is the result of stochastic tunneling through a metastable transfer potential. In accordance with reality, both wealth and income distribution transiently show Pareto tails for high income subjects. For metastable transfer potentials, exponential wealth condensation is a robust feature. For example with 10 % annual profit 1% of the population owns 50 % of the wealth after 50 years. The time to reach such a strong wealth condensation is a hyperbolic function of the annual profit rate.

**Introduction.** Modelling Economies focusses on deterministic agents in equilibrium. From the viewpoint of statistical mechanics, two essentials are missing: randomness and non-equilibrium dynamics. Randomness since agents decide from a chaotic, subjective environment. Since economies are built from exchange and credit relations between people and therefore between living disequilibrium entities, approaches far from equilibrium should be seriously pursued. Even if a steady state can be reached in principle, settling times are probably more on the order of years than days with persistent drifts in state variables.

Recent studies revealed exponential distributions in income data [1]-[3]. These findings were generalized to Boltzmann distributions with arbitrarily shaped thermodynamic potentials, describing economies similar to statistical thermodynamics [4]. The history and success of statistical mechanics should convince us that valuable insight can be gained from assuming that random fluctuations dominate agents in their behavior.

In taking a thermodynamic perspective, we describe economic agents on a thermodynamic basis and do not include all of their microeconomic details. Thermodynamics is successful, since energetic, not kinetic details of the interactions describe the system sufficiently on the macroscopic scale. We approach economics with the same spirit and argue that the analog of energy are transfer potentials [4]. The transfer characteristic of agents are derivatives of the transfer potential. This leads to a Boltzmann distribution of wealth with the transfer potential as energy term [4].

In following this idea, we focus on the energetic core of agent interaction. The ideal gas can be described so easily, because the energetics of the stochastic equilibrium matters. The microscopic modelling of the paths of every particle and the detailed dynamics of the collisions are not needed to find a first order approximation.

Influential studies discussed agent models of wealth accumulation and have clarified microeconomic details [5][6][7][8][9][10]. Here we focus in extracting the thermodynamic basic features that lead to strong condensation of wealth. In doing so, we approach economies from a global viewpoint with classical non-equilibrium thermodynamics. We start from basic economic transfers and discuss classes of economies which never reach equilibrium as they continuously spill out of a metastable transfer potentials.

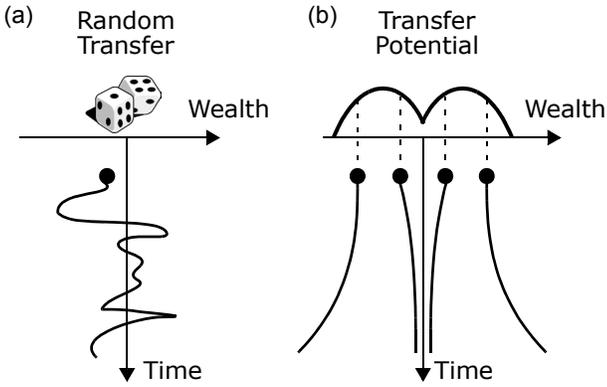
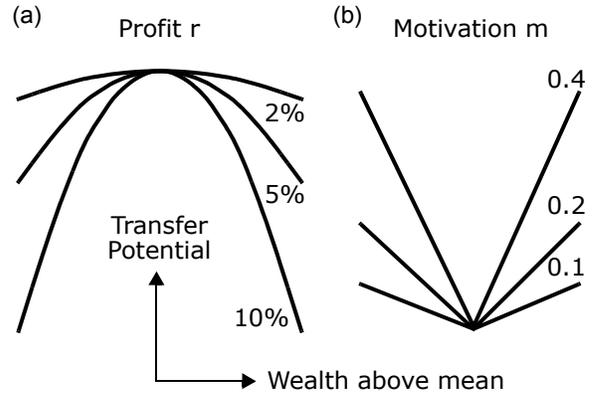

Fig. 1. **Basic Stochastic Economy.** Agents exchange wealth analogous to particles exchanging momentum. Following statistical mechanics, we divide the exchange into two classes. (a) Random wealth transfer, stemming from uncontrolled external variables are subsumed into a random transfer between agents. (b) Systematic wealth transfer between agents is subsumed into a transfer potential. The wealth gain per unit time of an agent is the derivative of the potential. As a result, agents ride down the hills of the potential.

Fig. 2. **Transfer potentials of the model economy: profit and Motivation.** We assume that two systematic transfers dominate the economy:. (a) Return of investment is proportional to the invested wealth. The profit rate r models the average gain from above-average wealth. Since wealth transfer is a zero-sum game, gains above average are balanced by losses below average. The result is a parabolic transfer potential. (b) Motivation is given by the constant gain or loss m depending on whether your wealth is below or above average. For simplicity, we model a single threshold of motivation. The resulting transfer potential stabilizes the economy in a local minima.

We have to note that we can only study relative wealth. Absolute wealth is notoriously difficult to define since it is never fully transferred between agents and therefore lacks monetary evaluation. However the dynamics of becoming relatively poor or relatively rich is visible on the market and can be monetarily evaluated. Relative wealth can be expected to however affect absolute wealth within a fast time scale.

**Economic Thermodynamics** (Fig. 1). The abstraction level of statistical mechanics is applied to agents in an economy. Thermalization between N agents is provided by N transfers of fixed wealth units $\Delta p$ per time step $\Delta t$ between randomly chosen agents (Fig. 1a). This defines a temperature given in units of diffusion constant $D = \Delta p^2 / \Delta t$. Systematic transfers F between agents are inferred from an external transfer potential U(p) with $F = -\nabla U$. For the case of a stable potential U, the wealth of the agents converge to a Boltzmann distribution $n(p) \propto \exp[-U(p)/D]$ with $\langle n \rangle = N$ [4].

This means that in statistical economies, the final wealth distribution can be directly inferred from the external transfers. Real world examples are societies which are dominated by social welfare systems and taxation [4]. Above approach is motivated by book-keeping: its relative wealth transfer mechanism was shown to have the same structure as momentum transfer in physics [4][11]. Notably, the potential U can describe a systematic bias of the random transfer process. For example debt limitations imposed on the random transfer can be modelled by an additional transfer potential [4].

**Profit and motivation yield a metastable economy** (Fig. 2). We extract three key factors of modern economies. First, incomplete information and chaotic environment results in a residual random transfers with diffusion constant D. Second, investments yield income with an annual profit rate r, thus leading to a harmonic transfer potential $U = -rp^2/2$ with linear income $F = -\nabla U = rp$ (Fig. 2a). The third income of an agent is due to motivation and depends on its relative wealth. Agents that have less-than-average wealth work harder than agents with above-average wealth. Without loss of generality, we choose a simple motivation: given the motivation parameter m, above average agents loose m per time step, below average gain m. The corresponding transfer potential is linear with $U = m|p - \bar{p}|$ and income $F = m\,\text{sign}(p - \bar{p})$ (Fig. 2b).

All three transfers mean that we describe an economy by thermal particles in the metastable potential given by (Fig. 3):

$$U(p) = -rp^2/2 + m|p - \bar{p}| \tag{1}$$

We see that the motivation parameter m breaks the symmetry of a harmonic profit economy. As easily seen from Eq. (1), profit rate r determines whether the economy diverges (r > 0) or converges (r < 0) into a steady state [4]. Focusing on profit-based economies, we study the diverging case with positive profit rate. The economy is characterized by a metastable transfer potential U since motivation gives a local stabilizing minima on a diverging global profit hill (Fig. 2, Fig. 3).

**Results.** We focus on cumulative wealth and income distributions. Cumulative wealth distribution $m(n_R)$ defines the percentage of wealth owned by the percentage $n_R$ of richest agents. It is calculated from the relative wealth distribution $n(p)$ by:

$$m(n_R) = \frac{W(p_0)}{W(-\infty)}$$

$$W(p_0) = \int_{p_0}^{\infty} |p| n(p) dp \tag{2}$$

$$n_R = \int_{p_0}^{\infty} n(p) dp$$

Similarly, a cumulative income distribution will show how much % of the total income is obtained by for example $n_I = 1\%$ of the agents with highest income. Typically we will monitor how much of total income or wealth is in the hands of 1 % agents with highest wealth or income. If more than 50 % of income or wealth is in the hands of 1 % of these top agents, we have a strong income or wealth accumulation.

**Wealth distribution dynamics of a metastable economy** (Fig. 3). We show the wealth dynamics of a typical metastable economy. We chose a wealth with profit rate r = 10 % / year with a motivation of m = 1 currency unit. As result, the transfer potential is metastable (Fig. 3). In a parabolic profit potential

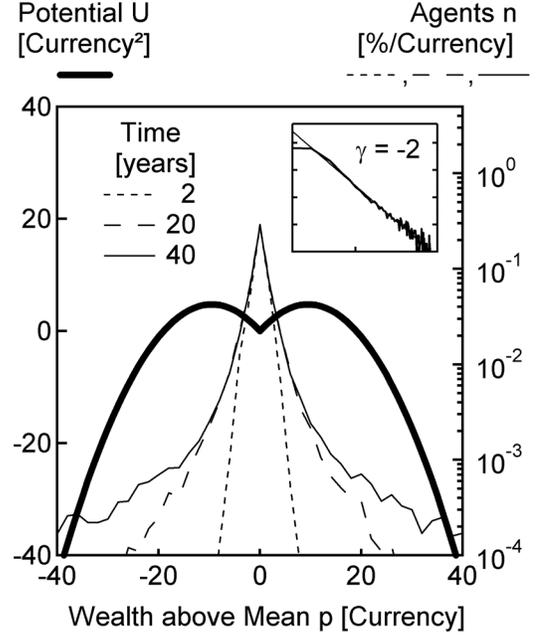

Fig. 3. **Wealth dynamics.** The combination of profit rate and motivation results in a parabolic profit potential with a central local motivation minima. Most of the agents are stabilized in the central motivation minimum. The potential however can be tunneled statistically, which leads to a minority of agents with fast growing wealth based on an unperturbed profit potential. Their distribution follows transiently at t=40Δt a Pareto power law (inset).

we find a central dip stemming from a local motivation minimum.

We start with all agents having identical mean wealth. This is a robust initial condition, starting with arbitrary distributions within the motivation minima - in our case ±10 currency units - leads to nearly identical results since most of the agents are stabilized in the central motivation minimum. In this local minima, agents fluctuate randomly in their relative wealth due to random wealth exchange.

For sufficient strong motivation, agents do not to overcome the central motivation dip. However, if they are lucky (or they work hard), they tunnel the potential statistically towards higher relative wealth p. Similarly, if they are less lucky (or work less), the agents drop out towards less relative wealth. Note that the profit parabola models both profit (e.g. interest of savings) and negative profit (e.g. interest from loans). For simplicity we start with a symmetric potential, the asymmetric case is discussed later in Fig. 7.

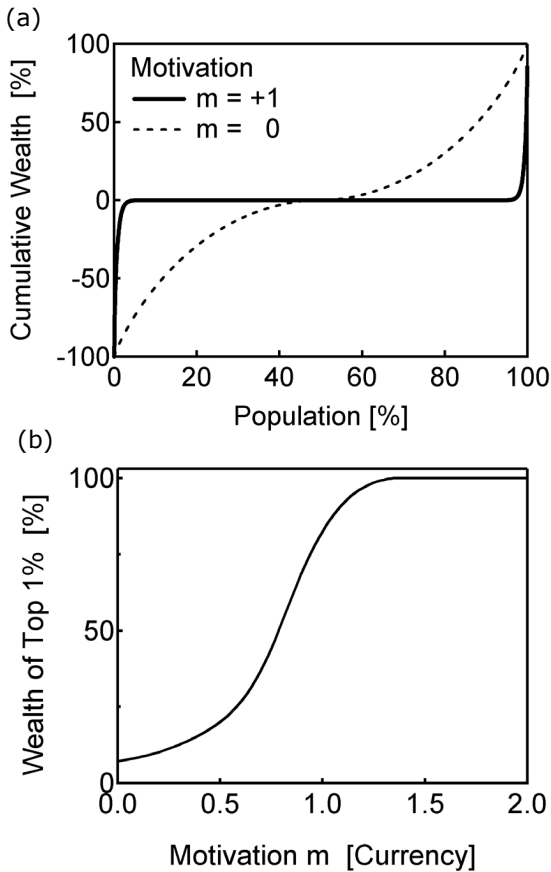

Fig. 4. **Strong Wealth Condensation.** (a) The cumulative wealth distribution shows strong condensation, visible at the sharp tails of the distribution. (b) Such strong wealth condensation is seen if the local motivation minimum is strong enough to keep most agents close to the average whereas only a minority can tunnel the potential to the undisturbed parabolic profit behavior. Typically this is seen for values of motivation m>1. In such cases, the top wealthy 1% own virtually 100% of above-average wealth. Notably, the final result does not depend on profit rate r.

The chosen symmetry is in principle nothing unusual. Credit money transfer between agents using bookkeeping is always a balance of given and received units of account, leading to the symmetric situation where the total of positive profits equal the total of negative profits [4].

When agents drop out of the central motivation dip due to high positive profit (p > 0) or high negative profit (p < 0, e.g. large investments), we find the identical dynamics. In both cases, the agents drift essentially deterministic towards positive or negative relative wealth. For sufficiently strong motivation m, only a minority tunnel the potential. For this wealthy minority, it is only a matter of time in the model to own 100 % of all wealth above mean.

For the symmetric case, the same applies to a poor minority dropping out into a debt trap towards p<0. Some people might argue that this is unrealistically since banks would not allow this. However even for asymmetric case with limited debt, identical wealth condensation is found on the rich side p>0 (Fig. 7).

Interestingly, the distribution of relative wealth of the rich agents follow transiently a Pareto power law [12] with exponent -2. This supports previous findings of Pareto exponents in real and simulated economies [5][6][12][13]. However the Pareto distribution is not stationary but only approximately valid around time point t =40 Δt. Wealth accumulation progresses and leads to further thickening of the fast tails as time progresses.

**Strong wealth condensation for sufficient motivation** (Fig. 4). Wealth condensation is best seen in the tails of the cumulative wealth distribution at infinite time (Fig. 4a). For sufficient motivation (m=1), strong wealth condensation is found: the distribution tails point sharply downward and upward at 0% and 100% of the wealth sorted population.

Note that for zero motivation (m=0), i.e. the case of a randomized, profit-only economy without threshold imposed by motivation, no significant wealth condensation is seen: the gaussian wealth distribution formed initially by the random process simply scales over time by the profit rate r. The economy freezes into an expansive steady state.

However, as motivation m rises, the tails of the distribution start to dominate. We plot total wealth of 1 % of top wealthy agents at infinite time versus the motivation m (Fig. 4b). It shows a sharp increase around m = 1, above which the economy converges to 1 % of top wealthy agents have virtually condensed all the wealth of the economy. It is important to understand that this behavior in the long run is insensitive to the value of the profit rate r chosen to be 10%/year for Fig. 4b.

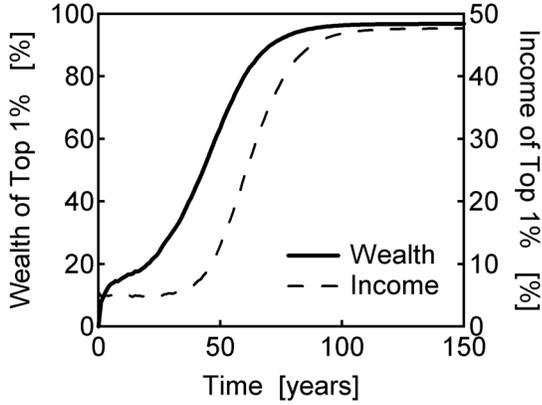

Fig. 5. **Dynamics of condensation: wealth leads, income follows.** The dynamics of wealth condensation is characterized by a slow creep phase, followed by an exponential condensation and a saturation phase leading to near 100% wealth accumulated in less than 1% of the agents. If there is no motivation (m=0) and the wealth dynamics is only governed by profit rate r, the system freezes into a steady state with gaussian wealth distribution and low wealth condensation. The condensation if seen in the cumulative income distribution with a similar dynamics. However the wealth condensation is seen in the income distribution only with a delay of about ten years for the shown profit rate of r = 10% / year.

**Dynamics of condensation: wealth leads, income follows** (Fig. 5). We follow the total wealth of the 1 % top wealthy agents over time for motivation m = 1 and profit rate r = 10 % / year. Wealth condensation shows fast dynamics over time. After a slow creep phase, nearly exponential condensation leads to a saturation near 100 %. Very interesting is that wealth condensation is ahead of income condensation. Only after wealth condensation is almost complete, signs of condensation in the income distribution is found. This seems counterintuitive, since one would expect that wealth accumulated income and therefore only after income shows fat distribution tails, wealth should show them. However, this is not the case.

From the point of view of potential tunneling from a metastable potential, it is clear that already small income advantages, not seen in income distribution, push agents over the threshold of the central motivation dip. This has important implications since in practice, inequality is often measured by income which is easy to collect from taxation records. However in our case, if one detects condensation in income, condensation of wealth is already far ahead and has reached serious levels.

**Wealth condensation is robust** (Fig. 6). Wealth condensation is robust and found for any profit rate r given that the motivation m is sufficiently large, typically above m>1. However, the speed of condensation for the wealth of top 1 % of agents depends strongly on profit rate r (Fig. 6a). The "half time" of condensation, i.e. the time to reach a 50% threshold of condensation is a hyperbolic function of profit rate r (Fig. 6b) given by

$$\tau_{50\%} = \frac{530\%}{r} \qquad (3)$$

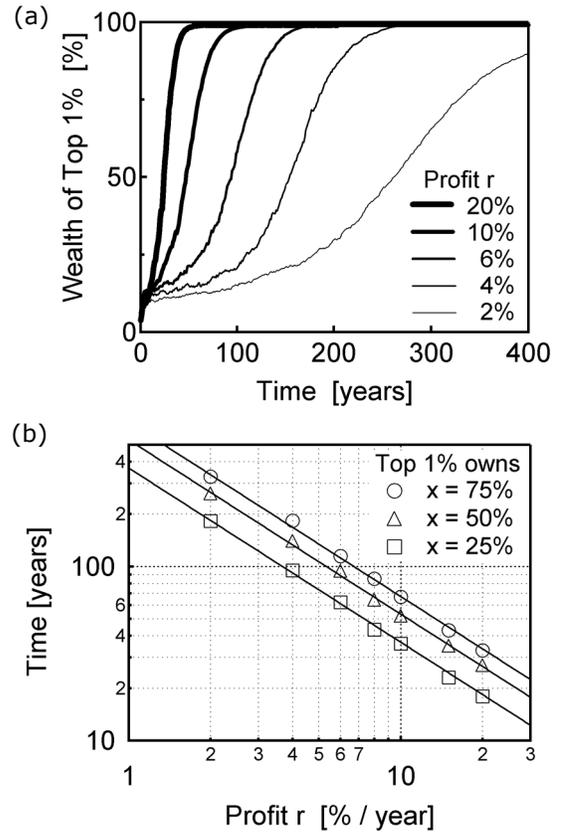

Fig. 6. **Wealth condensation is robust.** (a) The profit rate r only affects the time by which wealth condensation is complete. (b) For small profit rates of only r = 2% / year, it only takes about 265 years to leave more than 50% in the hands of the top 1% of agents. However for r = 20% / year the same condensation is already found after 26 years. This hyperbolic relationship holds for a number of condensation thresholds from 25% to 75%.

So for example, with a profit rate of r = 2% / year, the time to reach 50 % is 265 years. However, for a tenfold profit rate of 20 %, the same wealth condensation is reached after 26,5 years. Similar hyperbolic laws hold for a number of condensation thresholds (Fig. 6b).

**Wealth condensation is general feature of metastable potentials.** (Fig. 7) We have started for simplicity with a symmetric transfer potential. It might be correctly argued that agents with below-average wealth are typically not allowed be the credit market to step as deep into the debt trap as richer agents on the other side are gaining profits. Limitations of debt will lead to less fat tails in the distribution below the mean wealth.

The previous discussion however also applies for such conditions as shown for by a highly asymmetric example (Fig. 7a). The potential is still metastable due to a motivation dip, but a rising potential for agents with below-average wealth is implemented. The rise could be interpreted as social welfare transfers by the other agents, but it can be also understood as debt limitations by the credit market, imposing a bias potential for the random transfer. A similar debt limitation was discussed in [4], however for a converging stable potential.

Although the situation has totally changed for agents with negative relative wealth p, the cumulative distribution for p > 0 is essentially unchanged and both the dynamics and magnitude of wealth condensation is very similar to the previously discussed symmetric case. Again for a profit rate r = 10 %/year, top 1 % of above-average agents dominate 50 % of wealth after 50 years (Fig. 7b).

**Discussion.** We have approached wealth accumulation by non-equilibrium thermodynamics of random economies, modelled by the energy landscape of transfer potentials. We have demonstrated that a wide range of metastable transfer potentials are dangerous because they show a strong and fast dynamics of wealth condensation, interpreted physically as potential tunneling.

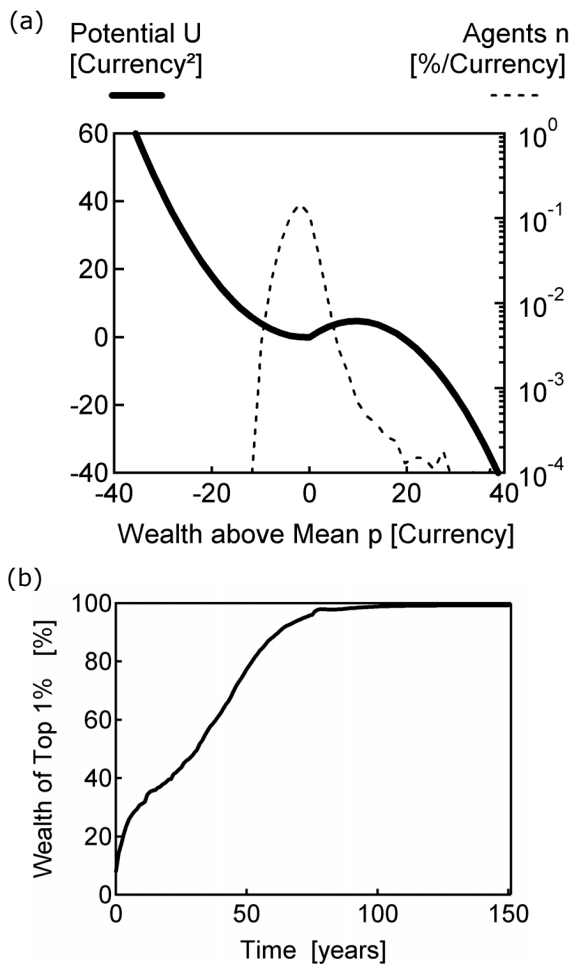

Fig. 7. **Wealth condensation is general feature of metastable transfer potentials.** Very similar wealth condensation is found for a completely different economy model. (a) Large loans are prohibited by a rising transfer potential for less-than-average wealth. (b) Yet the profit rate of 10% for above-mean wealth together with a motivation trough in the potential gives again a strong wealth condensation due to potential tunneling of the particles towards positive wealth.

So what are the arguments that modern economies are dominated by a metastable transfer landscape? We have argued in the beginning for the legitimacy of assuming profit rates, also known in less general terms as interest rates, together with motivational forces (Fig. 2). Why should economies and societies then start with a metastable transfer potential in the first place? What is the reason to construct metastable transfer potentials in an economy?

We gave microeconomical arguments, namely that below-average agents work harder and try to earn more than above-average agents, leading to a dent in the transfer potential (Fig. 2b, Fig. 3). However, metastability can also be interpreted on the global scale. One can argue that societies like to implement thresholds in their wealth dynamics to increase efficiency by motivation. And thresholds are best implemented with a local minima in the transfer potential for average wealth (p=0), leading to metastable potentials.

Metastability might result from the fact that basic consumption of average wealthy agents will always eat away part of the income, leaving less for profit-oriented investment. This argument does not apply for below-average agents, but the argument does not depend on symmetry as seen in Fig. 7. As a result, we again find a central dip in transfer potential and therefore to a metastable situation.

Note that it is hard to argue against the parabolic downward shape of the transfer potential for p>0. It is the result of the additivity of investment. The market generally makes no difference between one agent investing 2Δp and two agents investing Δp each. The direct result of this fact is the parabolic shape of the transfer potential which is the driving force of fast wealth growth at the distribution tails.

Notably, one of the ways to shape the transfer potential less steep for large p and therefore hindering the creation of fat distribution tails is non-flat progressive taxation, counteracting the additivity of investments. Similar converging potentials were discussed previously [4].

A clear result of the study is that wealth condensation critically depends on the depth of the metastable central dip. If this motivational dent at average wealth is too strong, creeping and strong wealth condensation is the result (Fig. 5, Fig. 6a, Fig. 7b). This dynamics is hard to detect since wealth does not follow income accumulation (Fig. 5), but wealth is concentrated before income.

To overcome condensation once it has established, it is of no use to make the transfer potential less metastable, since these changes for low |p| does not affect the distribution at the tails anymore. Melting the condensation would need changes in the profit structure for large |p|.

What we see is that motivation in an economy has two sides (Fig. 4b). If agents are not motivated (m=0), the economy freezes into an expanding state, probably leading to an inefficient economy. If however motivation is too high, we run into a creeping wealth condensation which is hard to detect and is exponentially self-energizing after a time given by the profit rate (Fig. 6). Whereas installation of motivation is probably fairly easy to establish, counteracting a first creeping, then running wealth condensation requires to change in the parabolic shape of the transfer potential and therefore strong interaction against free market principles.

**Conclusion (Fig. 8).** We have analyzed a statistical economy model of randomly exchanging agents under the influence of profit rate r and a motivational threshold m. We have discussed, how a physically motivated thermodynamic viewpoint give a global framework to describe economies. Motivation breaks the symmetry of the economy. For the most realistic case of positive profit, reasonable non-zero motivation leads to a fast diverging wealth condensation. For profit rates of 10 % / year, less than 50 years are needed to concentrate 50 % of the wealth into the hands of 1 % of the population.

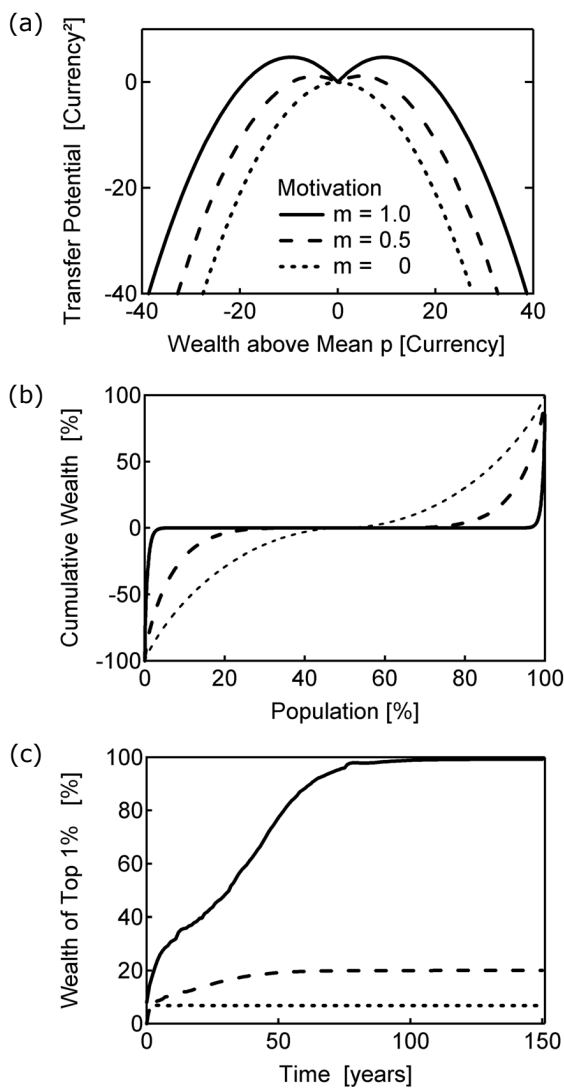

Fig. 8. **Conclusion.** Motivation m leads to an economy with a metastable transfer potential with a small central dip. It delicately decides between frozen economies (m = 0), moderate competition (m=0.5) or runaway wealth condensation (m>1).

The model shows that income distributions do not give adequate warning signs since their condensation lags the condensation of wealth. Moreover, the model generates transient Pareto distributions known from real world economies. Motivation has to be balanced (Fig. 8). Economies need motivation for innovation. However, if motivation is too strong, rare tunneling through metastable transfer potentials can be triggered, leading to strong wealth condensation. The thermodynamic viewpoint gives a simplified, global framework to monitor and control motivation without negative side effects.

We argue that the history of economies is paved with wealth condensation dynamics which start slow and often lead to social unrest. Understanding stabilizing factors on a global scale are crucial.

We thank Stephen Zarlenga, Benjamin Franksen and Robert Fischer for discussions.

1. Adrian Dragulescu and Victor M. Yakovenko, Statistical mechanics of money, European Physical Journal B, 17:723-729 (2000)
2. Adrian Dragulescu and Victor M. Yakovenko, Evidence for the exponential distribution of income in the USA, European Physical Journal B, 20:585-589 (2001)
3. Adrian Dragulescu and Victor M. Yakovenko, Exponential and power-law probability distributions of wealth and income in the United Kingdom and the United States, Physica A 299.213-221 (2001)
4. Robert Fischer and Dieter Braun, 2003, "Transfer Potentials shape and equilibrate Monetary Systems", Physica A 321:605-618 (2003).
5. Wealth condensation in a simple model of economy, Jean-Philippe Bouchaud, Marc Mézard, Physica A 282, 536-545 (2000)
6. Wealth condensation in pareto macroeconomies, Z. Burda, D. Johnston, J. Jurkiewicz, M. Kamin´ski, M. A. Nowak, G. Papp, and I. Zahed, Physical Review E 65:026102 (2002)
7. J. R. Iglesias, S. Gonçalves, S. Pianegonda, J. L. Vega and G. Abramson, Wealth redistribution in our small world, Physica A, 327:12-17 (2003)
8. N. Scafetta, S. Picozzi1 and B.J. West, An out-of-equilibrium model of the distributions of wealth, Quantitative Finance 4s:353-364 (2004)
9. Wealth accumulation with random redistribution, Ding-wei Huang, Physical Review E 69:057103 (2004)
10. Wealth distributions in asset exchange models, S. Ispolatov, P.L. Krapivsky, and S. Redner, Eur. Phys. J. B 2:267-276 (1998)
11. Robert Fischer and Dieter Braun, "Nontrivial Bookkeeping: a mechanical Perspective", Physica A 324:266-271 (2003)
12. V. Pareto, Cours d'économie politique. Reprinted as a volume of *Oeuvres Complètes* (Droz, Geneva, 1896-1965).
13. Z. Burda, D. Johnston, J. Jurkiewicz, M. Kamin´ski, M. A. Nowak, G. Papp and I. Zahed, Wealth condensation in pareto macroeconomies, Phys. Rev. E 65, 026102